\documentclass[aip,apl,reprint,twoside,floatfix,showpacs,showkeys]{revtex4-2}

\usepackage{amsmath,amssymb,amsfonts}
\usepackage{latexsym,eucal,gensymb}
\usepackage{dcolumn}
\usepackage{graphicx}
\usepackage{subfigure}
\usepackage{bm}
\usepackage[version=4]{mhchem}
\usepackage{miller}
\usepackage{textcomp}
\usepackage{array}
\usepackage{xspace}
\usepackage{color}
\usepackage{exscale}
\usepackage{xcolor}
\usepackage{physics}

\usepackage{ifpdf}
\ifpdf
\usepackage[pdftex,unicode,colorlinks,allcolors=blue,breaklinks,baseurl=http://]{hyperref}
\hypersetup{
	pdfauthor   = {Damir Islamov},%
	pdftitle    = {Impact of different metal electrodes on ferroelectric properties of hafnium-zirconium oxide},%
	pdfsubject  = {Ferroelectricity},%
	pdfkeywords = {hafnia, ferroelectrics, oxygen vacancy, charge transport, leakage current},%
	pdfcreator  = \LaTeX,%
	pdfproducer = \LaTeX}
\else
\usepackage[dvips,unicode,colorlinks,allcolors=blue,breaklinks,baseurl=http://]{hyperref}
\fi

\newcommand*{\T}[1]{\ensuremath{t_\mathrm{#1}}\xspace}
\newcommand*{\Fc}{\ensuremath{F_\mathrm{c}}\xspace}

\newcommand*{\Pres}{\ensuremath{P_\mathrm{r}}\xspace}
\newcommand*{\HZO}{\ce{Hf_{0.5}Zr_{0.5}O2}\xspace}

\begin{document}
\title{Changing the properties of Hf$_{0.5}$Zr$_{0.5}$O$_{2}$ during cyclic repolarization of ferroelectric capacitors with different electrode materials}

\author{Timur~M. Zalyalov}\email{timz@isp.nsc.ru}
\affiliation{Rzhanov Institute of Semiconductor Physics,
	Siberian Branch of the Russian Academy of Sciences,
	Novosibirsk 630090, Russian Federation}%
\affiliation{Novosibirsk State University,
	Novosibirsk 630090, Russian Federation}%
\author{Damir~R. Islamov}
\affiliation{Rzhanov Institute of Semiconductor Physics,
	Siberian Branch of the Russian Academy of Sciences,
	Novosibirsk 630090, Russian Federation}%
\affiliation{Novosibirsk State University,
	Novosibirsk 630090, Russian Federation}%

\date{\today}

\begin{abstract}
	The interest in the ferroelectric non-volatile memory
	as a candidate for low power consumption electronic memories
	was raised after the discovery of ferroelectricity in hafnium oxide.
	Doping by different elements of hafnia films allows improving
	their ferroelectric properties.
	In this work, the transport experiments are combined with the simulations
	to study the evolution of ferroelectric properties and
	the mean distance between oxygen vacancies
	during the endurance
	of hafnium-zirconium oxide in metal-ferroelectric-metal structures
	to study the impact of different metal electrodes.
\end{abstract}


\maketitle

The development of memory devices for using them in mobile systems, Internet-of-Things technology
and in other devices with high autonomy requires non-volatile
energy efficiency and high operation speed from used memory elements.
One of the candidates for the role of such memory is
Ferroelectric Random Access Memory (FRAM).
From the last decade of the last century to the present day,
FRAM elements used in microelectronics are based on lead zirconate-titanate (\ce{PZT}) \cite{Bondurant:FRAM:PZT:FM8x0y:1990}.
The problem of this material is the rapid disappearance of the ferroelectric response of a \ce{PZT} film
with a reduction in its thickness less than $100$\,nm.
This limitation made the FRAM production process technology to be unscalable beyond the $130$\,nm CMOS technology
(chips manufactured by Texas Instruments),
and, as a result, to a small information volume of integrated circuits.
Therefore, modern research in the field of FRAM technology is focused on studying
the ferroelectric properties of doped hafnium oxide \ce{HfO2} thin films.
The history of ferroelectric \ce{HfO2} began in $2007$, when the group members
at the Dynamic Random Access Memory (DRAM) company Qimonda,
together with the RWTH Aachen, searched
for materials for DRAM capacitor applications and
unexpectedly discovered the ferroelectric properties of doped thin \ce{HfO2} films.
It was demonstrated that the non-centrosymmetric polar orthorhombic phase (o-phase)
between the monoclinic (m-) and tetragonal (t-) phase grains
in thin ($5$--$50$\,nm) doped \ce{HfO2} layers
is responsible for maintaining polarization after the external electric field is turned off
\cite{FeRAM:HfO2:apl99:102903, o-phase:HfO2:APL99:112904, FE-HfO2:APL106:162905}.
The \ce{HfO2}-based FRAM technology demonstrates non-volatility,
high information volume per die,
high operation speed and is compatible with the modern CMOS technology \cite{28nm-FeFET:IEDM:2016}.

For more than $10$ years, researchers were working to solve problems
that prevent the implementation of \ce{HfO2}-based FRAM devices into mass-market production.
The main problems include a small memory window (MW$=2\Pres$, double value of the remnant polarization),
a small memory cell resource and instability of the memory window
during cyclic repolarization of ferroelectric.
A small memory window prevents a reliable reading logical state of the memory cell.
A small memory cell resource (endurance) is little information rewriting cycles
before the memory cell fails (breakdown).
The MW instability during the cycling consists in an increase in MW at the beginning
of cycling (wake-up effect)
and a decrease in the memory window before a breakdown (fatigue effect) \cite{Si-HfO2:FRAM:Mueller:DMR:2013, HfO2:FRAM:IRPS:2016}.
These problems can be solved by selecting doping \ce{HfO2} material
followed by rapid thermal annealing at high temperatures.
It was found that, the largest memory window is reached when using \ce{La} as a dopant \cite{HfO2:FRAM:Schroeder:2014}.
Also, when using \HZO instead of \ce{HfO2},
the wake-up effect is suppressed and the annealing temperature decreases \cite{HZO:FRAM:wakeup:Nanoscale:2016},
which is compatible with the thermal budget of back-end-of-line.
The ferroelectric response of the structure is also influenced by the method of metal contacts
deposition and the annealing temperature.
The highest memory window value for \HZO films without doping was obtained by applying
the \textsl{in vacuo} process, without atmospheric influence on the structure,
which made it possible to obtain a memory window of $2\Pres = 54.2$\,\textmu C/cm$^2$
due to the improvement of the ferroelectric/metal interface quality \cite{fe-HfZrO:largePr:apl:2021}.
Recently, it has been discovered that when \HZO is doped with \ce{La}, \ce{Y} and their mixture,
the films remain functional after $10^{11}$ rewriting cycles,
demonstrating quite a large MW value \cite{HZO:La:1e11endurance:jap:2019, fe-HfZrO:La-Y-Gd:1e11endurance:aem:2022}.
However, doping with \ce{La} led to a more pronounced manifestation of the wake-up effect.

The possible reason of this effect might be caused by the influence of metal electrodes.
The \HZO:\ce{La}-based structures in Refs.~\citenum{HZO:La:1e11endurance:jap:2019, fe-HfZrO:La-Y-Gd:1e11endurance:aem:2022}
had both bottom (BE) and top (TE) \ce{TiN} metal electrodes.
Recently, it has been shown that using \ce{RuO2} as the TE
in \ce{TiN}/\HZO/\ce{RuO2} \cite{TiN-HZO-TiN_RuO2:FRAM:PSS:2014},
as both the TE and the BE in \ce{RuO2}/\HZO/\ce{RuO2} \cite{fe-HfZrO:RuO2:jap:2021}
and \ce{Ru} as the BE in \ce{Ru}/\HZO/\ce{TiN} \cite{TiN_Ru-HZO-TiN:FRAM:apl:2020} structures
leads to a small memory window value and a relatively early breakdown during the electric field cycling.
Other metal electrodes were explored to promote the ferroelectric phase
in \ce{Hf_{1-x}Zr_{x}O2}
including \ce{TaN} \cite{HZO:TaN:jjap:2018, HfZrO-pyroelectric:apl:2017},
\ce{Ir} \cite{f-HfZrO-Ir:apl:2014, f-HfZrO-Ir:jjap:2014},
\ce{Pt} \cite{FeRAM:HfZrO:apl102:112914, f-HZO-Pt:apl104:072901, VO:FRAM:HfZrO:Shimizu:2015} and
\ce{W} \cite{f-HfZrO-W:apl:2017}
without endurance measurements.

In this letter, we investigate the impact of different metal electrodes
of \HZO in metal-ferroelectric-metal structures
during the endurance cycling.

\begin{table}
	\caption{Structures of all samples with different BE and TE under study}
	\label{t:samples}
	\centering
	\begin{tabular}{lccc}\hline\hline
		Sample	& Structure 				& BE 		& TE\\\hline
		S1		& \ce{TiN}/\HZO/\ce{TiN}	& \ce{TiN}	& \ce{TiN}\\
		S2		& \ce{TiN}/\HZO/\ce{RuO_x}	& \ce{TiN}	& \ce{RuO_x}\\
		S3		& \ce{TiN}/\HZO/\ce{TiAlN}	& \ce{TiN}	& \ce{TiAlN}\\
		S4		& \ce{TiN}/\HZO/\ce{NbN}	& \ce{TiN}	& \ce{NbN}\\
		S5		& \ce{W}/\HZO/\ce{TiN}		& \ce{W}	& \ce{TiN}\\
		S6		& \ce{W}/\HZO/\ce{W}		& \ce{W}	& \ce{W}\\
		S7		& \ce{TiN}/\HZO/\ce{W}		& \ce{TiN}	& \ce{W}\\
		\hline\hline
	\end{tabular}
\end{table}

\begin{figure}[tb]
	\includegraphics[width=0.6\columnwidth]{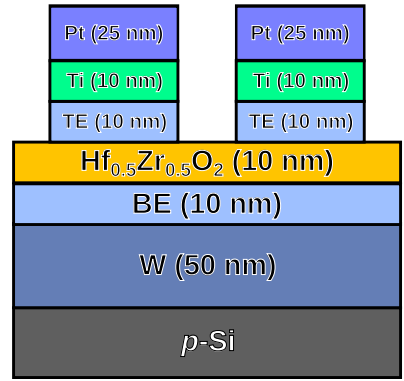}
	\caption{Schematic structure of the samples under study
		with different BE and TE.
	}
	\label{f:structures}
\end{figure}

The structures under study in this work were metal-1/\HZO/metal-2 (M1/FE/M2)
ferroelectric capacitors on silicon substrates
with the ferroelectric thicknesses of $10$\,nm.
Totally, seven kinds of samples with different electrodes were studied.
A description of all samples is provided in Table~\ref{t:samples}.
M1/FE/M2 capacitors were fabricated by the following procedure.
The $50$\,nm\,thick \ce{W} layer,
as well as $10$\,nm\,thick TE and BE were deposited on a silicon substrate
at room temperature using a sputtering system from Alliance Concept.
To produce S5 and S6 samples, the BE deposition step was skipped,
\textsl{i.e.}
\HZO films for these structures were deposited directly on the \ce{W} layers.
The $10$\,nm thick \HZO films were deposited by
atomic layer deposition (ALD) at deposition temperature $280$\celsius\xspace
in the Oxford Instruments OPAL system.
For hafnium and zirconium oxides,
tetrakis[ethylmethylamino]hafnium (\ce{Hf[N(C2H5)CH3]4}) and
tris[dimethylamino]cyclopentadienyl-zirconium (\ce{C5H5)Zr[N(CH3)2]3})
were used as precursors, respectively.
\ce{H2O} was used as an oxidant for all ALD cycles.
The samples were then treated by RTA at $600$\celsius\xspace for $20$\,s in the \ce{N2} atmosphere.
To pattern the capacitor structures,
a $10$\,nm thick \ce{Ti} layer was deposited as an adhesion layer,
and then $25$\,nm\,thick \ce{Pt} was deposited by the electron beam
evaporation through a shadow mask.
The round-shaped metal contact diameter was $200$\,\textmu{}m.
The wet chemical etching of TE was carried out for patterning using
the SC1 etching
($5\%$ \ce{NH3} and $2\%$ \ce{H2O2} solutions in \ce{H2O} at the temperature of $50$\celsius).
The schematic structures of the samples under study with different BE and TE
are shown in Fig.~\ref{f:structures}.
To produce S2, S6 and S7 samples, \ce{RuO2} or \ce{W} electrodes were deposited by physical vapor deposition
through a shadow mask on the \HZO films instead of the upper \ce{TiN} electrode.

\begin{figure}[tb]
	\includegraphics[width=\linewidth]{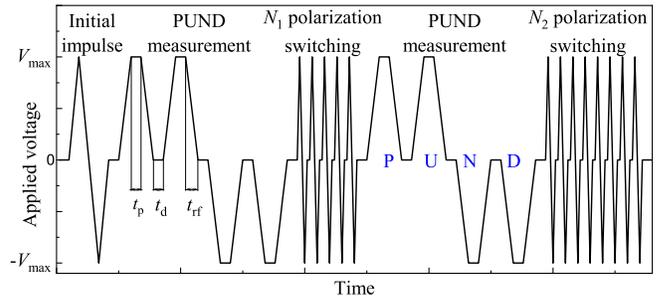}
	\caption{Schematic voltage sweeps during the endurance measurements at different points.
		Letters in blue show P, U, N and D pulses of the PUND sequence.}
	\label{f:pulses}
\end{figure}

An experimental study of the ferroelectric properties evolution during the structure state switching
was carried out by the cyclic application of trapezoidal voltage pulses of positive and negative amplitudes.
On some pulses, polarization-voltage ($P$-$V$) characteristics were measured by the PUND technique
followed by the DC current-voltage ($I$-$V$) measurements with the same voltage amplitude.
The measurements by the PUND method consist in the sequential application of two positive pulses (Positive/Up) and two negative (Negative/Down) to the structure.
The PUND sequences and cyclic voltage pulses are shown in Fig.~\ref{f:pulses} schematically.
The charge-carrier polarization values $\Pres$ during the cycling were obtained from the PUND measurements.
For the PUND measurements, the voltage rise/fall times $\T{rf}$, pulse delay $\T{d}$
and plateau time $\T{p}$ were the same and amounted $100$\,\textmu{}s.
The times for the cycling pulses were $\T{rf}=0.5$\,\textmu{}s, $\T{p}=2$\,\textmu{}s and $\T{d}=1.0$\,\textmu{}s.
The cycling pulse durations ensure that the polarization is switched to at least $80\%$ of the maximum possible value.
The voltage of all pulses was $3.0$\.V,
since at this amplitude value a compromise ratio of the memory window
and the cycling duration was observed~\cite{HfZrO-La:EDM:2022}.
The DC measurements were carried out from $-0.2$ to $+3.0$\,V
in $0.1$\,V increments for $7$\,sec.
Such long voltage sweep duration allows us
to reduce the displacement current contributions and
to extract their values at zero voltage.
The measurements were carried out using a Keithley 4200-SCS parametric analyzer
with 4225-PMU (pulses and PUND) and SMU-4210 (DC) units at room temperature.
The experiments were carried out up to the sample break down
or when $\Pres$ becomes negligible.

Following the PUND method, DC measurements made it possible to obtain leakage currents.
As a result of averaging the currents measured during the forward ($0 \to \pm V$) and reverse ($\pm V \to 0$) voltage sweep, displacement currents were excluded, and leakage currents were obtained:
\begin{equation}
	I_\textrm{leakage}(V) = {1}/{2}\left(I^\textrm{DC}_{0 \to +V}(V)+I^\textrm{DC}_{+V \to 0}(V)\right).
	\label{e:leakage}
\end{equation}

\begin{figure}[tb]
	\includegraphics[width=\columnwidth]{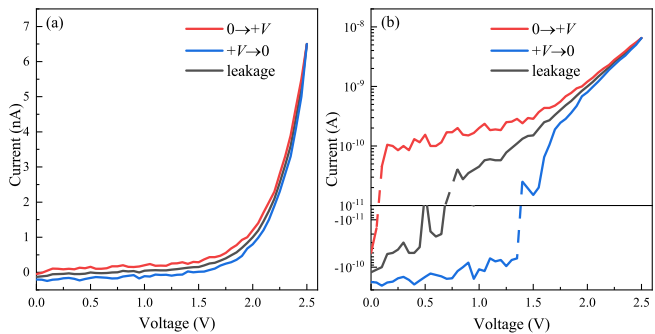}
	\subfigure{\label{f:leakage:lin}}
	\subfigure{\label{f:leakage:log}}
	\caption{\subref{f:leakage:lin} Currents flowing when the voltage increases $I_{0 \to +V}(V)$
	    (red lines) and when the voltage decreases $I_{+V \to 0}(V)$ (blue line).
	    The displacement currents $\pm C(V) dV/dt$ change sign as the voltage increases and decreases, while the leakage currents $I_\textrm{leakage}(V)$ do not.
	    As a result of averaging the currents of $I_{0 \to +V}(V)$ and the $I_{+V \to 0}(V)$, only leakage currents (black lines) remain.
		\subref{f:leakage:log} The same data in the semilog plate.
		Dashed lines connect positive and negative current values.
	}
	\label{f:leakage}
\end{figure}

The procedure for extracting the leakage current is illustrated in Fig.~\ref{f:leakage}.
The mean distance between neighboring traps $a$ was extracted from the $I$-$V$ dependences
of the leakage currents
within the model of phonon-assisted tunneling of electrons (holes) between neighboring traps
(PATENT) \cite{Si3N4:transport:PAT:JETP139:1026:en}.
Recently, it has been demonstrated that this transport model
adequately describes the leakages in both amorphous and ferroelectric \ce{HfO2}-based films
\cite{HfO2:Transport:2014,defects-HfO2:PhysRep:2016,transport:HfO2La:APL:2020}
as well as in \HZO-based structures \cite{HfZrO:transport:2015, a+f-HfZrO:transport:2015:en,HZO:ActMater:2019}.
This procedure is described in details elsewhere \cite{HfO2+ZrO2:Autom2017en,HZO:ActMater:2019,drift+diffusion-trapped-charge:SciRep:2020}.

\begin{figure}[!tb]
	\includegraphics[width=\columnwidth]{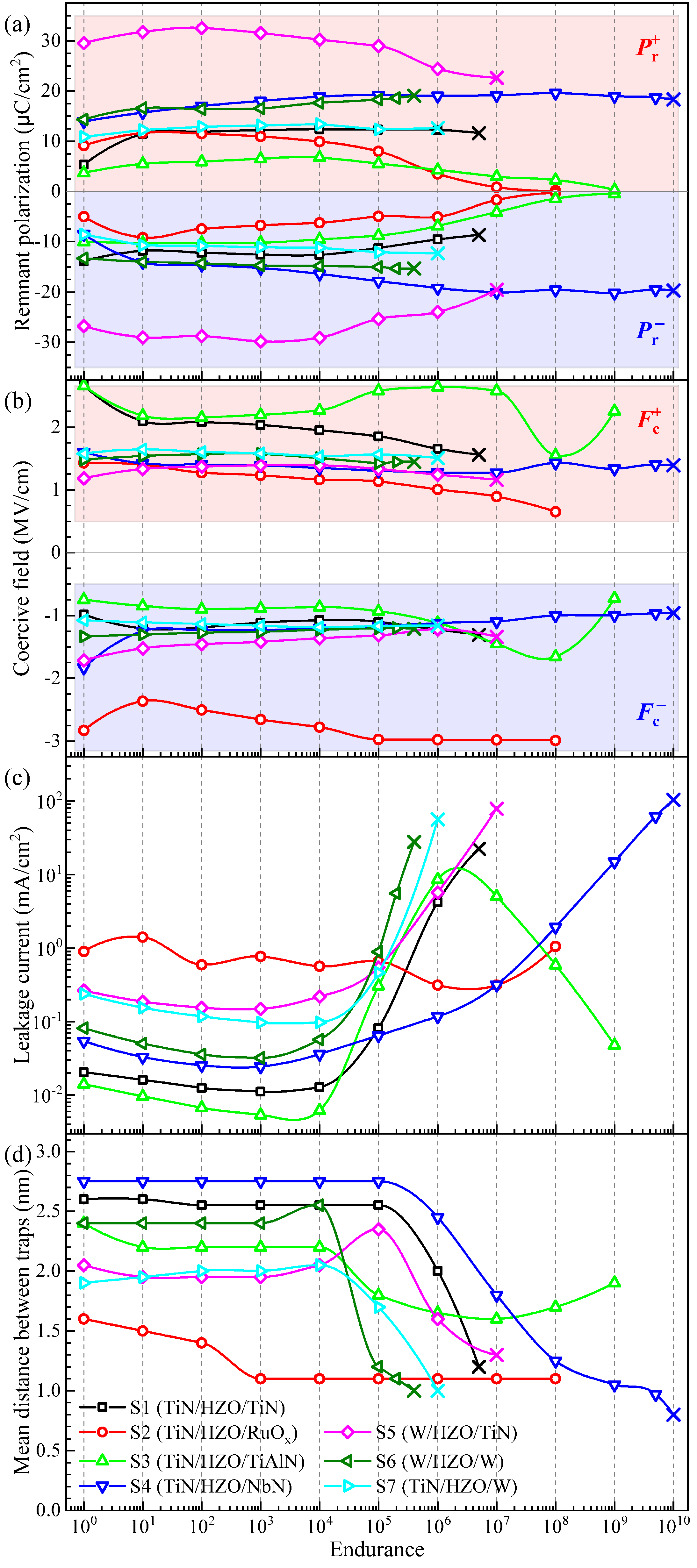}
	\subfigure{\label{f:endurance:MW}}
	\subfigure{\label{f:endurance:Fc}}
	\subfigure{\label{f:endurance:Ileak}}
	\subfigure{\label{f:endurance:a}}
	\caption{Evolution of the ferroelectric and transport properties
		on the M1/FE/M2 structures with different bottom and top electrodes:
		\subref{f:endurance:MW}~ferroelectric memory windows,
		\subref{f:endurance:Fc}~coercive fields,
		\subref{f:endurance:Ileak}~the leakage current values at $3.0$\,V and
		\subref{f:endurance:a}~the mean distance between traps.
		Crosses show the last obtained data before breakdown.
		Vertical dashed lines are visual guides.}
	\label{f:endurance}
\end{figure}

The presence of \HZO film ferroelectric properties is confirmed by observing
the hysteresis in the po\-la\-ri\-za\-ti\-on-vol\-tage characteristics
for all M1/FE/M2 structures.
In Fig.~\ref{f:endurance} is the evolution of the ferroelectric and transport properties
of M1/FE/M2 structures with different bottom and top metal electrodes.
The S1 sample with \ce{TiN} BE and TE exhibits a typical charge-carrier polarization ($\Pres^\pm$) evolution
(black characters in Fig.~\ref{f:endurance:MW}) in accordance with literature data \cite{FeRAM:HfZrO:apl99:112901, HZO:ActMater:2019}:
it endures more than $10^6$ switch cycles with a slight wake-up and fatigue effects
with $\mathrm{MW}=|\Pres^+|+|\Pres^-|\approx 20$\,\textmu C/cm$^2$
and then breaks down, as shown in Fig.~\ref{f:endurance} by crosses.
The $\Pres^\pm$ values for S2 and S3 samples are less than $10$\,\textmu C/cm$^2$
and disappear after $10^{8}$--$10^{9}$ switching cycles.
The S5 sample exhibits the largest residual polarization
with $\mathrm{MW} \approx 60$\,\textmu C/cm$^2$.
After expressive fatigue effect, the sample reaches a breakdown after $10^{7}$ endurance cycles.
The MW values for samples S1, S4, S6 and S7 are close,
but all of them, except S4, are broken down after $10^5$--$10^7$ switch cycles.
The sample with \ce{NbN} TE (S4) endures more than $10^{10}$ repolarization
with $\mathrm{MW} \approx 30$--$40$\,\textmu C/cm$^2$.

The coercive field values $\Fc^\pm$ (Fig.~\ref{f:endurance:Fc}) are constant with small deviations
for almost all structures under study except for S2 and S3 samples.
As far as the MW disappears, $\Fc^\pm$ loses its physical meaning.
One can see that measured $\Fc^\pm$ values exhibit slight asymmetry.
This can be caused by built-in electric field due to
different work functions of the metal electrodes.
Also, asymmetric interfaces around ferroelectric films,
due to the exposure of all BEs to the ALD processing temperature,
can introduce distortions in the $\Fc^\pm$ measurements.

The leakage currents, measured at the voltage of $3$\,V are shown in Fig.~\ref{f:endurance:Ileak}.
One can see that leakages rise after $10^4$ cycles for all structures, but the S2 sample,
which exhibits constant leakage current.
The leakage current of structure S3 shows complex evolution:
namely, up to $10^4$ endurance cycles, its value decreases,
then increases,
and after $10^6$ cycles it decreases to the original value.

The extracted
mean distances between neighboring traps
are shown in Fig.~\ref{f:endurance:a}.
One can see that the distances between traps in ferroelectrics of almost all structures
remain at the same levels up to $10^4$ cycles of repolarization.
Reducing the average distances between traps after $10^4$--$10^5$ switching cycles
are accompanied by an increase in leakage currents
due to their exponential dependence~\cite{Si3N4:transport:PAT:JETP139:1026:en}.

Since the traps in \HZO are oxygen vacancies,
a decrease in the distances between traps
is associated with new oxygen vacancy generation \cite{HfZrO:transport:2015, HZO:ActMater:2019}.
Oxygen ions that left the crystal lattice can migrate in an external electric field
and even leave the \HZO film for a metal electrode \cite{ox-vac:migration:science:2021}.
The S4 structure is formed with the top \ce{NbN} electrode,
\textsl{i.e.}, it is more corrosion and oxidation resistant than \ce{TiN} \cite{TiN+NbN:corrosion:TSF:2004}.
Thus, the \ce{TiN} electrode in the structures absorbs more oxygen than \ce{NbN}.
This effect leads to a greater endurance of the structure during repolarization
even with a large number of defects.

It is difficult to determine where new oxygen vacancies are located.
We assume that, initially, there are many traps in the grain boundaries
and much less in the grain bulk.
Thus, the leakage current is caused by traps at the grain boundaries.
After $10^4$ cycles new traps are generated in the grain bulk due to the oxygen ion migration
during the cyclic switching,
and some ions \ce{O-} might reach grain boundaries or even the \HZO/metal interface
followed by the absorption of electrodes.
Then the leakage currents are limited by the traps in the grain bulk.
In this case, part of the applied voltage drops at the interface layer between \HZO and the electrode,
such as \ce{WO_x} for \ce{W}~\cite{M1-HZO-M2:FRAM:AFM:2023:2303261}
or \ce{TiON} for \ce{TiN}~\cite{FRAM:Gd:HfO2:AElM:2016}.
If the traps are located in the grain boundaries,
then the mean distance between the traps $a$ can be converted to the surface (2D) trap density
$N_\mathrm{2D} = a^{-2} \sim 10^{13}$--$10^{14}$\,cm$^{-2}$.
If the traps are distributed in the grain bulk,
then the volumetric (3D) trap density is
$N_\mathrm{3D} = a^{-3} \sim 10^{19}$--$10^{21}$\,cm$^{-3}$.

It is interesting to note that the S5 structure
with \ce{W} BE exhibits an extremely high MW value, compared to other structures.
We assume that this is caused by the absence of the interface sublayer \ce{WO_x}
with high static dielectric constant
after the \HZO deposition.
This is confirmed by the HAXPES analysis conducted
for the same set of structures~\cite{M1-HZO-M2:FRAM:AFM:2023:2303261}.
The endurance cycling leads to a fast generation of new oxygen vacancies in \HZO,
forming the sub-layer followed by a breakdown.
\ce{RuO_x} and \ce{TiAlN} electrodes absorb oxygen more actively than other metals,
the interface sublayer is formed faster, and the voltage drops on it \cite{HZO+IL:IMW:2022}.
This leads to the vanishing of the ferroelectric response in \HZO films
and decrease in leakage currents.
It should be noted that these samples were not broken down.
One can assume that \ce{RuO_x} and \ce{TiAlN} electrodes are good for the DRAM application.

Also, it is interesting to note that
S5 and S7 structures are symmetrical to each other,
but exhibit their different ferroelectric evolution during the endurance procedure.
We assume that this is caused by different temperature influences
on the BE ($280$\textcelsius) and the TE (room temperature)
during the ferroelectric film deposition and formation the structure as a whole.
It should be noted that not only the electrode material,
but also its specific properties such as grain texture
can affect the \HZO film properties and performance.
The study of the metal electrodes structural properties
and their effect on the \HZO film ferroelectric properties
is beyond the scope of this work and is a task for the future.

In conclusion, transport experiments, combined with simulations,
were studied for thin \HZO ferroelectric film-based structures with different metal electrodes.
It was found that \ce{RuO_x} and \ce{TiAlN} electrodes suppress the ferroelectric response
in \HZO films, as well as leakages through the structure.
The utilization of \ce{NbN} as an electrode in M/FE/M structures allows obtaining a more stable
ferroelectric capacitor with more than $10^{10}$ cycle switches.

\begin{acknowledgments}
The work was performed within the framework of the state assignment of the Ministry of Education and Science
of the Russian Federation for the Rzhanov Institute of Semiconductor Physics of
the Siberian Branch of the Russian Academy of Sciences.
The authors would like to express their gratitude to
Prof.~Thomas Mikolajick,
Dr.~Uwe Schroeder,
Ruben Alcala and
Monica Materano for the provided samples.
The authors would like to acknowledge and thank
Dr.~Uwe Schroeder and
Dr.~Alexey Mironov
for their valuable discussions.
\end{acknowledgments}

\section*{Conflict of Interest}
The authors have no conflicts to disclose.

\section*{Data Availability}
The data that support the findings of this study are available
from the corresponding author upon reasonable request.

\bibliographystyle{apsrev4-2}
\bibliography{IEEEabrv,../../../../bibtex}

 \providecommand{\noop}[1]{}
\begin{thebibliography}{39}%
\makeatletter
\providecommand \@ifxundefined [1]{%
 \@ifx{#1\undefined}
}%
\providecommand \@ifnum [1]{%
 \ifnum #1\expandafter \@firstoftwo
 \else \expandafter \@secondoftwo
 \fi
}%
\providecommand \@ifx [1]{%
 \ifx #1\expandafter \@firstoftwo
 \else \expandafter \@secondoftwo
 \fi
}%
\providecommand \natexlab [1]{#1}%
\providecommand \enquote  [1]{``#1''}%
\providecommand \bibnamefont  [1]{#1}%
\providecommand \bibfnamefont [1]{#1}%
\providecommand \citenamefont [1]{#1}%
\providecommand \href@noop [0]{\@secondoftwo}%
\providecommand \href [0]{\begingroup \@sanitize@url \@href}%
\providecommand \@href[1]{\@@startlink{#1}\@@href}%
\providecommand \@@href[1]{\endgroup#1\@@endlink}%
\providecommand \@sanitize@url [0]{\catcode `\\12\catcode `\$12\catcode
  `\&12\catcode `\#12\catcode `\^12\catcode `\_12\catcode `\%12\relax}%
\providecommand \@@startlink[1]{}%
\providecommand \@@endlink[0]{}%
\providecommand \url  [0]{\begingroup\@sanitize@url \@url }%
\providecommand \@url [1]{\endgroup\@href {#1}{\urlprefix }}%
\providecommand \urlprefix  [0]{URL }%
\providecommand \Eprint [0]{\href }%
\providecommand \doibase [0]{https://doi.org/}%
\providecommand \selectlanguage [0]{\@gobble}%
\providecommand \bibinfo  [0]{\@secondoftwo}%
\providecommand \bibfield  [0]{\@secondoftwo}%
\providecommand \translation [1]{[#1]}%
\providecommand \BibitemOpen [0]{}%
\providecommand \bibitemStop [0]{}%
\providecommand \bibitemNoStop [0]{.\EOS\space}%
\providecommand \EOS [0]{\spacefactor3000\relax}%
\providecommand \BibitemShut  [1]{\csname bibitem#1\endcsname}%
\let\auto@bib@innerbib\@empty
\bibitem [{\citenamefont {Bondurant}(1990)}]{Bondurant:FRAM:PZT:FM8x0y:1990}%
  \BibitemOpen
  \bibfield  {author} {\bibinfo {author} {\bibfnamefont {D.}~\bibnamefont
  {Bondurant}},\ }\href {https://doi.org/10.1080/00150199008008233} {\bibfield
  {journal} {\bibinfo  {journal} {Ferroelectrics}\ }\textbf {\bibinfo {volume}
  {112}},\ \bibinfo {pages} {273} (\bibinfo {year} {1990})}\BibitemShut
  {NoStop}%
\bibitem [{\citenamefont {B\"{o}scke}\ \emph
  {et~al.}(2011{\natexlab{a}})\citenamefont {B\"{o}scke}, \citenamefont
  {M\"{u}ller}, \citenamefont {Br\"{a}uhaus}, \citenamefont {Schr\"{o}der},\
  and\ \citenamefont {B\"{o}ttger}}]{FeRAM:HfO2:apl99:102903}%
  \BibitemOpen
  \bibfield  {author} {\bibinfo {author} {\bibfnamefont {T.~S.}\ \bibnamefont
  {B\"{o}scke}}, \bibinfo {author} {\bibfnamefont {J.}~\bibnamefont
  {M\"{u}ller}}, \bibinfo {author} {\bibfnamefont {D.}~\bibnamefont
  {Br\"{a}uhaus}}, \bibinfo {author} {\bibfnamefont {U.}~\bibnamefont
  {Schr\"{o}der}},\ and\ \bibinfo {author} {\bibfnamefont {U.}~\bibnamefont
  {B\"{o}ttger}},\ }\href {https://doi.org/10.1063/1.3634052} {\bibfield
  {journal} {\bibinfo  {journal} {Applied Physics Letters}\ }\textbf {\bibinfo
  {volume} {99}},\ \bibinfo {eid} {102903} (\bibinfo {year}
  {2011}{\natexlab{a}})}\BibitemShut {NoStop}%
\bibitem [{\citenamefont {B\"{o}scke}\ \emph
  {et~al.}(2011{\natexlab{b}})\citenamefont {B\"{o}scke}, \citenamefont
  {Teichert}, \citenamefont {Br\"{a}uhaus}, \citenamefont {M\"{u}ller},
  \citenamefont {Schr\"{o}der}, \citenamefont {B\"{o}ttger},\ and\
  \citenamefont {Mikolajick}}]{o-phase:HfO2:APL99:112904}%
  \BibitemOpen
  \bibfield  {author} {\bibinfo {author} {\bibfnamefont {T.~S.}\ \bibnamefont
  {B\"{o}scke}}, \bibinfo {author} {\bibfnamefont {S.}~\bibnamefont
  {Teichert}}, \bibinfo {author} {\bibfnamefont {D.}~\bibnamefont
  {Br\"{a}uhaus}}, \bibinfo {author} {\bibfnamefont {J.}~\bibnamefont
  {M\"{u}ller}}, \bibinfo {author} {\bibfnamefont {U.}~\bibnamefont
  {Schr\"{o}der}}, \bibinfo {author} {\bibfnamefont {U.}~\bibnamefont
  {B\"{o}ttger}},\ and\ \bibinfo {author} {\bibfnamefont {T.}~\bibnamefont
  {Mikolajick}},\ }\href {https://doi.org/10.1063/1.3636434} {\bibfield
  {journal} {\bibinfo  {journal} {Applied Physics Letters}\ }\textbf {\bibinfo
  {volume} {99}},\ \bibinfo {pages} {112904} (\bibinfo {year}
  {2011}{\natexlab{b}})}\BibitemShut {NoStop}%
\bibitem [{\citenamefont {Sang}\ \emph {et~al.}(2015)\citenamefont {Sang},
  \citenamefont {Grimley}, \citenamefont {Schenk}, \citenamefont {Schroeder},\
  and\ \citenamefont {LeBeau}}]{FE-HfO2:APL106:162905}%
  \BibitemOpen
  \bibfield  {author} {\bibinfo {author} {\bibfnamefont {X.}~\bibnamefont
  {Sang}}, \bibinfo {author} {\bibfnamefont {E.~D.}\ \bibnamefont {Grimley}},
  \bibinfo {author} {\bibfnamefont {T.}~\bibnamefont {Schenk}}, \bibinfo
  {author} {\bibfnamefont {U.}~\bibnamefont {Schroeder}},\ and\ \bibinfo
  {author} {\bibfnamefont {J.~M.}\ \bibnamefont {LeBeau}},\ }\href
  {https://doi.org/10.1063/1.4919135} {\bibfield  {journal} {\bibinfo
  {journal} {Applied Physics Letters}\ }\textbf {\bibinfo {volume} {106}},\
  \bibinfo {eid} {162905} (\bibinfo {year} {2015})}\BibitemShut {NoStop}%
\bibitem [{\citenamefont {Trentzsch}\ \emph {et~al.}(2016)\citenamefont
  {Trentzsch}, \citenamefont {Flachowsky}, \citenamefont {Richter},
  \citenamefont {Paul}, \citenamefont {Reimer}, \citenamefont {Utess},
  \citenamefont {Jansen}, \citenamefont {Mulaosmanovic}, \citenamefont
  {Muller}, \citenamefont {Slesazeck}, \citenamefont {Ocker}, \citenamefont
  {Noack}, \citenamefont {Muller}, \citenamefont {Polakowski}, \citenamefont
  {Schreiter}, \citenamefont {Beyer}, \citenamefont {Mikolajick},\ and\
  \citenamefont {Rice}}]{28nm-FeFET:IEDM:2016}%
  \BibitemOpen
  \bibfield  {author} {\bibinfo {author} {\bibfnamefont {M.}~\bibnamefont
  {Trentzsch}}, \bibinfo {author} {\bibfnamefont {S.}~\bibnamefont
  {Flachowsky}}, \bibinfo {author} {\bibfnamefont {R.}~\bibnamefont {Richter}},
  \bibinfo {author} {\bibfnamefont {J.}~\bibnamefont {Paul}}, \bibinfo {author}
  {\bibfnamefont {B.}~\bibnamefont {Reimer}}, \bibinfo {author} {\bibfnamefont
  {D.}~\bibnamefont {Utess}}, \bibinfo {author} {\bibfnamefont
  {S.}~\bibnamefont {Jansen}}, \bibinfo {author} {\bibfnamefont
  {H.}~\bibnamefont {Mulaosmanovic}}, \bibinfo {author} {\bibfnamefont
  {S.}~\bibnamefont {Muller}}, \bibinfo {author} {\bibfnamefont
  {S.}~\bibnamefont {Slesazeck}}, \bibinfo {author} {\bibfnamefont
  {J.}~\bibnamefont {Ocker}}, \bibinfo {author} {\bibfnamefont
  {M.}~\bibnamefont {Noack}}, \bibinfo {author} {\bibfnamefont
  {J.}~\bibnamefont {Muller}}, \bibinfo {author} {\bibfnamefont
  {P.}~\bibnamefont {Polakowski}}, \bibinfo {author} {\bibfnamefont
  {J.}~\bibnamefont {Schreiter}}, \bibinfo {author} {\bibfnamefont
  {S.}~\bibnamefont {Beyer}}, \bibinfo {author} {\bibfnamefont
  {T.}~\bibnamefont {Mikolajick}},\ and\ \bibinfo {author} {\bibfnamefont
  {B.}~\bibnamefont {Rice}},\ }in\ \href
  {https://doi.org/10.1109/iedm.2016.7838397} {\emph {\bibinfo {booktitle}
  {2016 {IEEE} International Electron Devices Meeting ({IEDM})}}}\ (\bibinfo
  {publisher} {{IEEE}},\ \bibinfo {year} {2016})\ pp.\ \bibinfo {pages}
  {11.5.1--11.5.4}\BibitemShut {NoStop}%
\bibitem [{\citenamefont {Mueller}\ \emph {et~al.}(2013)\citenamefont
  {Mueller}, \citenamefont {Muller}, \citenamefont {Schroeder},\ and\
  \citenamefont {Mikolajick}}]{Si-HfO2:FRAM:Mueller:DMR:2013}%
  \BibitemOpen
  \bibfield  {author} {\bibinfo {author} {\bibfnamefont {S.}~\bibnamefont
  {Mueller}}, \bibinfo {author} {\bibfnamefont {J.}~\bibnamefont {Muller}},
  \bibinfo {author} {\bibfnamefont {U.}~\bibnamefont {Schroeder}},\ and\
  \bibinfo {author} {\bibfnamefont {T.}~\bibnamefont {Mikolajick}},\ }\href
  {https://doi.org/10.1109/tdmr.2012.2216269} {\bibfield  {journal} {\bibinfo
  {journal} {{IEEE} Trans. Device Mater. Rel.}\ }\textbf {\bibinfo {volume}
  {13}},\ \bibinfo {pages} {93} (\bibinfo {year} {2013})}\BibitemShut {NoStop}%
\bibitem [{\citenamefont {Pesi\v{c}}\ \emph {et~al.}(2016)\citenamefont
  {Pesi\v{c}}, \citenamefont {Fengler}, \citenamefont {Slesazeck},
  \citenamefont {Schroeder}, \citenamefont {Mikolajick}, \citenamefont
  {Larcher},\ and\ \citenamefont {Padovani}}]{HfO2:FRAM:IRPS:2016}%
  \BibitemOpen
  \bibfield  {author} {\bibinfo {author} {\bibfnamefont {M.}~\bibnamefont
  {Pesi\v{c}}}, \bibinfo {author} {\bibfnamefont {F.~P.~G.}\ \bibnamefont
  {Fengler}}, \bibinfo {author} {\bibfnamefont {S.}~\bibnamefont {Slesazeck}},
  \bibinfo {author} {\bibfnamefont {U.}~\bibnamefont {Schroeder}}, \bibinfo
  {author} {\bibfnamefont {T.}~\bibnamefont {Mikolajick}}, \bibinfo {author}
  {\bibfnamefont {L.}~\bibnamefont {Larcher}},\ and\ \bibinfo {author}
  {\bibfnamefont {A.}~\bibnamefont {Padovani}},\ }in\ \href
  {https://doi.org/10.1109/irps.2016.7574619} {\emph {\bibinfo {booktitle}
  {2016 {IEEE} International Reliability Physics Symposium ({IRPS})}}}\
  (\bibinfo  {publisher} {{IEEE}},\ \bibinfo {year} {2016})\ pp.\ \bibinfo
  {pages} {MY--3--1--MY--3--5}\BibitemShut {NoStop}%
\bibitem [{\citenamefont {Schroeder}\ \emph {et~al.}(2014)\citenamefont
  {Schroeder}, \citenamefont {Yurchuk}, \citenamefont {M\"{u}ller},
  \citenamefont {Martin}, \citenamefont {Schenk}, \citenamefont {Polakowski},
  \citenamefont {Adelmann}, \citenamefont {Popovici}, \citenamefont {Kalinin},\
  and\ \citenamefont {Mikolajick}}]{HfO2:FRAM:Schroeder:2014}%
  \BibitemOpen
  \bibfield  {author} {\bibinfo {author} {\bibfnamefont {U.}~\bibnamefont
  {Schroeder}}, \bibinfo {author} {\bibfnamefont {E.}~\bibnamefont {Yurchuk}},
  \bibinfo {author} {\bibfnamefont {J.}~\bibnamefont {M\"{u}ller}}, \bibinfo
  {author} {\bibfnamefont {D.}~\bibnamefont {Martin}}, \bibinfo {author}
  {\bibfnamefont {T.}~\bibnamefont {Schenk}}, \bibinfo {author} {\bibfnamefont
  {P.}~\bibnamefont {Polakowski}}, \bibinfo {author} {\bibfnamefont
  {C.}~\bibnamefont {Adelmann}}, \bibinfo {author} {\bibfnamefont {M.~I.}\
  \bibnamefont {Popovici}}, \bibinfo {author} {\bibfnamefont {S.~V.}\
  \bibnamefont {Kalinin}},\ and\ \bibinfo {author} {\bibfnamefont
  {T.}~\bibnamefont {Mikolajick}},\ }\href
  {https://doi.org/10.7567/jjap.53.08le02} {\bibfield  {journal} {\bibinfo
  {journal} {Japanese Journal of Applied Physics}\ }\textbf {\bibinfo {volume}
  {53}},\ \bibinfo {pages} {08LE02} (\bibinfo {year} {2014})}\BibitemShut
  {NoStop}%
\bibitem [{\citenamefont {Kim}\ \emph {et~al.}(2016)\citenamefont {Kim},
  \citenamefont {Park}, \citenamefont {Kim}, \citenamefont {Lee}, \citenamefont
  {Moon}, \citenamefont {Kim}, \citenamefont {Hyun},\ and\ \citenamefont
  {Hwang}}]{HZO:FRAM:wakeup:Nanoscale:2016}%
  \BibitemOpen
  \bibfield  {author} {\bibinfo {author} {\bibfnamefont {H.~J.}\ \bibnamefont
  {Kim}}, \bibinfo {author} {\bibfnamefont {M.~H.}\ \bibnamefont {Park}},
  \bibinfo {author} {\bibfnamefont {Y.~J.}\ \bibnamefont {Kim}}, \bibinfo
  {author} {\bibfnamefont {Y.~H.}\ \bibnamefont {Lee}}, \bibinfo {author}
  {\bibfnamefont {T.}~\bibnamefont {Moon}}, \bibinfo {author} {\bibfnamefont
  {K.~D.}\ \bibnamefont {Kim}}, \bibinfo {author} {\bibfnamefont {S.~D.}\
  \bibnamefont {Hyun}},\ and\ \bibinfo {author} {\bibfnamefont {C.~S.}\
  \bibnamefont {Hwang}},\ }\href {https://doi.org/10.1039/c5nr05339k}
  {\bibfield  {journal} {\bibinfo  {journal} {Nanoscale}\ }\textbf {\bibinfo
  {volume} {8}},\ \bibinfo {pages} {1383} (\bibinfo {year} {2016})}\BibitemShut
  {NoStop}%
\bibitem [{\citenamefont {Lee}\ \emph {et~al.}(2021)\citenamefont {Lee},
  \citenamefont {Alex~Hsain}, \citenamefont {Fields}, \citenamefont
  {Jaszewski}, \citenamefont {Horgan}, \citenamefont {Edgington}, \citenamefont
  {Ihlefeld}, \citenamefont {Parsons},\ and\ \citenamefont
  {Jones}}]{fe-HfZrO:largePr:apl:2021}%
  \BibitemOpen
  \bibfield  {author} {\bibinfo {author} {\bibfnamefont {Y.}~\bibnamefont
  {Lee}}, \bibinfo {author} {\bibfnamefont {H.}~\bibnamefont {Alex~Hsain}},
  \bibinfo {author} {\bibfnamefont {S.~S.}\ \bibnamefont {Fields}}, \bibinfo
  {author} {\bibfnamefont {S.~T.}\ \bibnamefont {Jaszewski}}, \bibinfo {author}
  {\bibfnamefont {M.~D.}\ \bibnamefont {Horgan}}, \bibinfo {author}
  {\bibfnamefont {P.~G.}\ \bibnamefont {Edgington}}, \bibinfo {author}
  {\bibfnamefont {J.~F.}\ \bibnamefont {Ihlefeld}}, \bibinfo {author}
  {\bibfnamefont {G.~N.}\ \bibnamefont {Parsons}},\ and\ \bibinfo {author}
  {\bibfnamefont {J.~L.}\ \bibnamefont {Jones}},\ }\href
  {https://doi.org/10.1063/5.0029532} {\bibfield  {journal} {\bibinfo
  {journal} {Applied Physics Letters}\ }\textbf {\bibinfo {volume} {118}},\
  \bibinfo {eid} {012903} (\bibinfo {year} {2021})}\BibitemShut {NoStop}%
\bibitem [{\citenamefont {Kozodaev}\ \emph {et~al.}(2019)\citenamefont
  {Kozodaev}, \citenamefont {Chernikova}, \citenamefont {Korostylev},
  \citenamefont {Park}, \citenamefont {Khakimov}, \citenamefont {Hwang},\ and\
  \citenamefont {Markeev}}]{HZO:La:1e11endurance:jap:2019}%
  \BibitemOpen
  \bibfield  {author} {\bibinfo {author} {\bibfnamefont {M.~G.}\ \bibnamefont
  {Kozodaev}}, \bibinfo {author} {\bibfnamefont {A.~G.}\ \bibnamefont
  {Chernikova}}, \bibinfo {author} {\bibfnamefont {E.~V.}\ \bibnamefont
  {Korostylev}}, \bibinfo {author} {\bibfnamefont {M.~H.}\ \bibnamefont
  {Park}}, \bibinfo {author} {\bibfnamefont {R.~R.}\ \bibnamefont {Khakimov}},
  \bibinfo {author} {\bibfnamefont {C.~S.}\ \bibnamefont {Hwang}},\ and\
  \bibinfo {author} {\bibfnamefont {A.~M.}\ \bibnamefont {Markeev}},\ }\href
  {https://doi.org/10.1063/1.5050700} {\bibfield  {journal} {\bibinfo
  {journal} {Journal of Applied Physics}\ }\textbf {\bibinfo {volume} {125}},\
  \bibinfo {eid} {034101} (\bibinfo {year} {2019})}\BibitemShut {NoStop}%
\bibitem [{\citenamefont {Popovici}\ \emph {et~al.}(2022)\citenamefont
  {Popovici}, \citenamefont {Walke}, \citenamefont {Bizindavyi}, \citenamefont
  {Meersschaut}, \citenamefont {Banerjee}, \citenamefont {Potoms},
  \citenamefont {Katcko}, \citenamefont {Van~den Bosch}, \citenamefont
  {Delhougne}, \citenamefont {Kar},\ and\ \citenamefont
  {Van~Houdt}}]{fe-HfZrO:La-Y-Gd:1e11endurance:aem:2022}%
  \BibitemOpen
  \bibfield  {author} {\bibinfo {author} {\bibfnamefont {M.~I.}\ \bibnamefont
  {Popovici}}, \bibinfo {author} {\bibfnamefont {A.~M.}\ \bibnamefont {Walke}},
  \bibinfo {author} {\bibfnamefont {J.}~\bibnamefont {Bizindavyi}}, \bibinfo
  {author} {\bibfnamefont {J.}~\bibnamefont {Meersschaut}}, \bibinfo {author}
  {\bibfnamefont {K.}~\bibnamefont {Banerjee}}, \bibinfo {author}
  {\bibfnamefont {G.}~\bibnamefont {Potoms}}, \bibinfo {author} {\bibfnamefont
  {K.}~\bibnamefont {Katcko}}, \bibinfo {author} {\bibfnamefont
  {G.}~\bibnamefont {Van~den Bosch}}, \bibinfo {author} {\bibfnamefont
  {R.}~\bibnamefont {Delhougne}}, \bibinfo {author} {\bibfnamefont {G.~S.}\
  \bibnamefont {Kar}},\ and\ \bibinfo {author} {\bibfnamefont {J.}~\bibnamefont
  {Van~Houdt}},\ }\href {https://doi.org/10.1021/acsaelm.2c00063} {\bibfield
  {journal} {\bibinfo  {journal} {{ACS} Applied Electronic Materials}\ }\textbf
  {\bibinfo {volume} {4}},\ \bibinfo {pages} {1823} (\bibinfo {year}
  {2022})}\BibitemShut {NoStop}%
\bibitem [{\citenamefont {Park}\ \emph
  {et~al.}(2014{\natexlab{a}})\citenamefont {Park}, \citenamefont {Kim},
  \citenamefont {Kim}, \citenamefont {Jeon}, \citenamefont {Moon},\ and\
  \citenamefont {Hwang}}]{TiN-HZO-TiN_RuO2:FRAM:PSS:2014}%
  \BibitemOpen
  \bibfield  {author} {\bibinfo {author} {\bibfnamefont {M.~H.}\ \bibnamefont
  {Park}}, \bibinfo {author} {\bibfnamefont {H.~J.}\ \bibnamefont {Kim}},
  \bibinfo {author} {\bibfnamefont {Y.~J.}\ \bibnamefont {Kim}}, \bibinfo
  {author} {\bibfnamefont {W.}~\bibnamefont {Jeon}}, \bibinfo {author}
  {\bibfnamefont {T.}~\bibnamefont {Moon}},\ and\ \bibinfo {author}
  {\bibfnamefont {C.~S.}\ \bibnamefont {Hwang}},\ }\href
  {https://doi.org/10.1002/pssr.201409017} {\bibfield  {journal} {\bibinfo
  {journal} {physica status solidi (RRL) --- Rapid Research Letters}\ }\textbf
  {\bibinfo {volume} {8}},\ \bibinfo {pages} {532} (\bibinfo {year}
  {2014}{\natexlab{a}})}\BibitemShut {NoStop}%
\bibitem [{\citenamefont {Fields}\ \emph {et~al.}(2021)\citenamefont {Fields},
  \citenamefont {Smith}, \citenamefont {Jaszewski}, \citenamefont {Mimura},
  \citenamefont {Dickie}, \citenamefont {Esteves}, \citenamefont {David~Henry},
  \citenamefont {Wolfley}, \citenamefont {Davids},\ and\ \citenamefont
  {Ihlefeld}}]{fe-HfZrO:RuO2:jap:2021}%
  \BibitemOpen
  \bibfield  {author} {\bibinfo {author} {\bibfnamefont {S.~S.}\ \bibnamefont
  {Fields}}, \bibinfo {author} {\bibfnamefont {S.~W.}\ \bibnamefont {Smith}},
  \bibinfo {author} {\bibfnamefont {S.~T.}\ \bibnamefont {Jaszewski}}, \bibinfo
  {author} {\bibfnamefont {T.}~\bibnamefont {Mimura}}, \bibinfo {author}
  {\bibfnamefont {D.~A.}\ \bibnamefont {Dickie}}, \bibinfo {author}
  {\bibfnamefont {G.}~\bibnamefont {Esteves}}, \bibinfo {author} {\bibfnamefont
  {M.}~\bibnamefont {David~Henry}}, \bibinfo {author} {\bibfnamefont {S.~L.}\
  \bibnamefont {Wolfley}}, \bibinfo {author} {\bibfnamefont {P.~S.}\
  \bibnamefont {Davids}},\ and\ \bibinfo {author} {\bibfnamefont {J.~F.}\
  \bibnamefont {Ihlefeld}},\ }\href {https://doi.org/10.1063/5.0064145}
  {\bibfield  {journal} {\bibinfo  {journal} {Journal of Applied Physics}\
  }\textbf {\bibinfo {volume} {130}},\ \bibinfo {eid} {134101} (\bibinfo {year}
  {2021})}\BibitemShut {NoStop}%
\bibitem [{\citenamefont {Chernikova}\ \emph {et~al.}(2020)\citenamefont
  {Chernikova}, \citenamefont {Kozodaev}, \citenamefont {Khakimov},
  \citenamefont {Polyakov},\ and\ \citenamefont
  {Markeev}}]{TiN_Ru-HZO-TiN:FRAM:apl:2020}%
  \BibitemOpen
  \bibfield  {author} {\bibinfo {author} {\bibfnamefont {A.~G.}\ \bibnamefont
  {Chernikova}}, \bibinfo {author} {\bibfnamefont {M.~G.}\ \bibnamefont
  {Kozodaev}}, \bibinfo {author} {\bibfnamefont {R.~R.}\ \bibnamefont
  {Khakimov}}, \bibinfo {author} {\bibfnamefont {S.~N.}\ \bibnamefont
  {Polyakov}},\ and\ \bibinfo {author} {\bibfnamefont {A.~M.}\ \bibnamefont
  {Markeev}},\ }\href {https://doi.org/10.1063/5.0022118} {\bibfield  {journal}
  {\bibinfo  {journal} {Applied Physics Letters}\ }\textbf {\bibinfo {volume}
  {117}},\ \bibinfo {eid} {192902} (\bibinfo {year} {2020})}\BibitemShut
  {NoStop}%
\bibitem [{\citenamefont {Migita}\ \emph {et~al.}(2018)\citenamefont {Migita},
  \citenamefont {Ota}, \citenamefont {Yamada}, \citenamefont {Shibuya},
  \citenamefont {Sawa},\ and\ \citenamefont {Toriumi}}]{HZO:TaN:jjap:2018}%
  \BibitemOpen
  \bibfield  {author} {\bibinfo {author} {\bibfnamefont {S.}~\bibnamefont
  {Migita}}, \bibinfo {author} {\bibfnamefont {H.}~\bibnamefont {Ota}},
  \bibinfo {author} {\bibfnamefont {H.}~\bibnamefont {Yamada}}, \bibinfo
  {author} {\bibfnamefont {K.}~\bibnamefont {Shibuya}}, \bibinfo {author}
  {\bibfnamefont {A.}~\bibnamefont {Sawa}},\ and\ \bibinfo {author}
  {\bibfnamefont {A.}~\bibnamefont {Toriumi}},\ }\href
  {https://doi.org/10.7567/jjap.57.04fb01} {\bibfield  {journal} {\bibinfo
  {journal} {Japanese Journal of Applied Physics}\ }\textbf {\bibinfo {volume}
  {57}},\ \bibinfo {pages} {04FB01} (\bibinfo {year} {2018})}\BibitemShut
  {NoStop}%
\bibitem [{\citenamefont {Smith}\ \emph {et~al.}(2017)\citenamefont {Smith},
  \citenamefont {Kitahara}, \citenamefont {Rodriguez}, \citenamefont {Henry},
  \citenamefont {Brumbach},\ and\ \citenamefont
  {Ihlefeld}}]{HfZrO-pyroelectric:apl:2017}%
  \BibitemOpen
  \bibfield  {author} {\bibinfo {author} {\bibfnamefont {S.~W.}\ \bibnamefont
  {Smith}}, \bibinfo {author} {\bibfnamefont {A.~R.}\ \bibnamefont {Kitahara}},
  \bibinfo {author} {\bibfnamefont {M.~A.}\ \bibnamefont {Rodriguez}}, \bibinfo
  {author} {\bibfnamefont {M.~D.}\ \bibnamefont {Henry}}, \bibinfo {author}
  {\bibfnamefont {M.~T.}\ \bibnamefont {Brumbach}},\ and\ \bibinfo {author}
  {\bibfnamefont {J.~F.}\ \bibnamefont {Ihlefeld}},\ }\href
  {https://doi.org/10.1063/1.4976519} {\bibfield  {journal} {\bibinfo
  {journal} {Applied Physics Letters}\ }\textbf {\bibinfo {volume} {110}},\
  \bibinfo {eid} {072901} (\bibinfo {year} {2017})}\BibitemShut {NoStop}%
\bibitem [{\citenamefont {Park}\ \emph
  {et~al.}(2014{\natexlab{b}})\citenamefont {Park}, \citenamefont {Kim},
  \citenamefont {Kim}, \citenamefont {Lee}, \citenamefont {Moon}, \citenamefont
  {Kim},\ and\ \citenamefont {Hwang}}]{f-HfZrO-Ir:apl:2014}%
  \BibitemOpen
  \bibfield  {author} {\bibinfo {author} {\bibfnamefont {M.~H.}\ \bibnamefont
  {Park}}, \bibinfo {author} {\bibfnamefont {H.~J.}\ \bibnamefont {Kim}},
  \bibinfo {author} {\bibfnamefont {Y.~J.}\ \bibnamefont {Kim}}, \bibinfo
  {author} {\bibfnamefont {W.}~\bibnamefont {Lee}}, \bibinfo {author}
  {\bibfnamefont {T.}~\bibnamefont {Moon}}, \bibinfo {author} {\bibfnamefont
  {K.~D.}\ \bibnamefont {Kim}},\ and\ \bibinfo {author} {\bibfnamefont {C.~S.}\
  \bibnamefont {Hwang}},\ }\href {https://doi.org/10.1063/1.4893376} {\bibfield
   {journal} {\bibinfo  {journal} {Applied Physics Letters}\ }\textbf {\bibinfo
  {volume} {105}},\ \bibinfo {eid} {072902} (\bibinfo {year}
  {2014}{\natexlab{b}})}\BibitemShut {NoStop}%
\bibitem [{\citenamefont {Shimizu}\ \emph {et~al.}(2014)\citenamefont
  {Shimizu}, \citenamefont {Yokouchi}, \citenamefont {Shiraishi}, \citenamefont
  {Oikawa}, \citenamefont {Krishnan},\ and\ \citenamefont
  {Funakubo}}]{f-HfZrO-Ir:jjap:2014}%
  \BibitemOpen
  \bibfield  {author} {\bibinfo {author} {\bibfnamefont {T.}~\bibnamefont
  {Shimizu}}, \bibinfo {author} {\bibfnamefont {T.}~\bibnamefont {Yokouchi}},
  \bibinfo {author} {\bibfnamefont {T.}~\bibnamefont {Shiraishi}}, \bibinfo
  {author} {\bibfnamefont {T.}~\bibnamefont {Oikawa}}, \bibinfo {author}
  {\bibfnamefont {P.~S. S.~R.}\ \bibnamefont {Krishnan}},\ and\ \bibinfo
  {author} {\bibfnamefont {H.}~\bibnamefont {Funakubo}},\ }\href
  {https://doi.org/10.7567/jjap.53.09pa04} {\bibfield  {journal} {\bibinfo
  {journal} {Japanese Journal of Applied Physics}\ }\textbf {\bibinfo {volume}
  {53}},\ \bibinfo {pages} {09PA04} (\bibinfo {year} {2014})}\BibitemShut
  {NoStop}%
\bibitem [{\citenamefont {Park}\ \emph {et~al.}(2013)\citenamefont {Park},
  \citenamefont {Kim}, \citenamefont {Kim}, \citenamefont {Lee}, \citenamefont
  {Kim},\ and\ \citenamefont {Hwang}}]{FeRAM:HfZrO:apl102:112914}%
  \BibitemOpen
  \bibfield  {author} {\bibinfo {author} {\bibfnamefont {M.~H.}\ \bibnamefont
  {Park}}, \bibinfo {author} {\bibfnamefont {H.~J.}\ \bibnamefont {Kim}},
  \bibinfo {author} {\bibfnamefont {Y.~J.}\ \bibnamefont {Kim}}, \bibinfo
  {author} {\bibfnamefont {W.}~\bibnamefont {Lee}}, \bibinfo {author}
  {\bibfnamefont {H.~K.}\ \bibnamefont {Kim}},\ and\ \bibinfo {author}
  {\bibfnamefont {C.~S.}\ \bibnamefont {Hwang}},\ }\href
  {https://doi.org/10.1063/1.4798265} {\bibfield  {journal} {\bibinfo
  {journal} {Applied Physics Letters}\ }\textbf {\bibinfo {volume} {102}},\
  \bibinfo {pages} {112914} (\bibinfo {year} {2013})}\BibitemShut {NoStop}%
\bibitem [{\citenamefont {Park}\ \emph
  {et~al.}(2014{\natexlab{c}})\citenamefont {Park}, \citenamefont {Kim},
  \citenamefont {Kim}, \citenamefont {Moon},\ and\ \citenamefont
  {Hwang}}]{f-HZO-Pt:apl104:072901}%
  \BibitemOpen
  \bibfield  {author} {\bibinfo {author} {\bibfnamefont {M.~H.}\ \bibnamefont
  {Park}}, \bibinfo {author} {\bibfnamefont {H.~J.}\ \bibnamefont {Kim}},
  \bibinfo {author} {\bibfnamefont {Y.~J.}\ \bibnamefont {Kim}}, \bibinfo
  {author} {\bibfnamefont {T.}~\bibnamefont {Moon}},\ and\ \bibinfo {author}
  {\bibfnamefont {C.~S.}\ \bibnamefont {Hwang}},\ }\href
  {https://doi.org/10.1063/1.4866008} {\bibfield  {journal} {\bibinfo
  {journal} {Applied Physics Letters}\ }\textbf {\bibinfo {volume} {104}},\
  \bibinfo {eid} {072901} (\bibinfo {year} {2014}{\natexlab{c}})}\BibitemShut
  {NoStop}%
\bibitem [{\citenamefont {Shimizu}\ \emph {et~al.}(2015)\citenamefont
  {Shimizu}, \citenamefont {Yokouchi}, \citenamefont {Oikawa}, \citenamefont
  {Shiraishi}, \citenamefont {Kiguchi}, \citenamefont {Akama}, \citenamefont
  {Konno}, \citenamefont {Gruverman},\ and\ \citenamefont
  {Funakubo}}]{VO:FRAM:HfZrO:Shimizu:2015}%
  \BibitemOpen
  \bibfield  {author} {\bibinfo {author} {\bibfnamefont {T.}~\bibnamefont
  {Shimizu}}, \bibinfo {author} {\bibfnamefont {T.}~\bibnamefont {Yokouchi}},
  \bibinfo {author} {\bibfnamefont {T.}~\bibnamefont {Oikawa}}, \bibinfo
  {author} {\bibfnamefont {T.}~\bibnamefont {Shiraishi}}, \bibinfo {author}
  {\bibfnamefont {T.}~\bibnamefont {Kiguchi}}, \bibinfo {author} {\bibfnamefont
  {A.}~\bibnamefont {Akama}}, \bibinfo {author} {\bibfnamefont {T.~J.}\
  \bibnamefont {Konno}}, \bibinfo {author} {\bibfnamefont {A.}~\bibnamefont
  {Gruverman}},\ and\ \bibinfo {author} {\bibfnamefont {H.}~\bibnamefont
  {Funakubo}},\ }\href {https://doi.org/10.1063/1.4915336} {\bibfield
  {journal} {\bibinfo  {journal} {Applied Physics Letters}\ }\textbf {\bibinfo
  {volume} {106}},\ \bibinfo {eid} {112904} (\bibinfo {year}
  {2015})}\BibitemShut {NoStop}%
\bibitem [{\citenamefont {Karbasian}\ \emph {et~al.}(2017)\citenamefont
  {Karbasian}, \citenamefont {dos Reis}, \citenamefont {Yadav}, \citenamefont
  {Tan}, \citenamefont {Hu},\ and\ \citenamefont
  {Salahuddin}}]{f-HfZrO-W:apl:2017}%
  \BibitemOpen
  \bibfield  {author} {\bibinfo {author} {\bibfnamefont {G.}~\bibnamefont
  {Karbasian}}, \bibinfo {author} {\bibfnamefont {R.}~\bibnamefont {dos Reis}},
  \bibinfo {author} {\bibfnamefont {A.~K.}\ \bibnamefont {Yadav}}, \bibinfo
  {author} {\bibfnamefont {A.~J.}\ \bibnamefont {Tan}}, \bibinfo {author}
  {\bibfnamefont {C.}~\bibnamefont {Hu}},\ and\ \bibinfo {author}
  {\bibfnamefont {S.}~\bibnamefont {Salahuddin}},\ }\href
  {https://doi.org/10.1063/1.4993739} {\bibfield  {journal} {\bibinfo
  {journal} {Applied Physics Letters}\ }\textbf {\bibinfo {volume} {111}},\
  \bibinfo {eid} {022907} (\bibinfo {year} {2017})}\BibitemShut {NoStop}%
\bibitem [{\citenamefont {Zalyalov}\ and\ \citenamefont
  {Islamov}(2022)}]{HfZrO-La:EDM:2022}%
  \BibitemOpen
  \bibfield  {author} {\bibinfo {author} {\bibfnamefont {T.~M.}\ \bibnamefont
  {Zalyalov}}\ and\ \bibinfo {author} {\bibfnamefont {D.~R.}\ \bibnamefont
  {Islamov}},\ }in\ \href {https://doi.org/10.1109/edm55285.2022.9855130}
  {{\emph {\bibinfo {booktitle} {2022 {IEEE} 23rd
  International Conference of Young Professionals in Electron Devices and
  Materials ({EDM})}}}},\ \bibinfo {organization} {IEEE}\ (\bibinfo
  {publisher} {{IEEE}},\ \bibinfo {year} {2022})\ pp.\ \bibinfo {pages}
  {48--51}\BibitemShut {NoStop}%
\bibitem [{\citenamefont {Nasyrov}\ and\ \citenamefont
  {Gritsenko}(2011)}]{Si3N4:transport:PAT:JETP139:1026:en}%
  \BibitemOpen
  \bibfield  {author} {\bibinfo {author} {\bibfnamefont {K.~A.}\ \bibnamefont
  {Nasyrov}}\ and\ \bibinfo {author} {\bibfnamefont {V.~A.}\ \bibnamefont
  {Gritsenko}},\ }\href {https://doi.org/10.1134/S1063776111040200} {\bibfield
  {journal} {\bibinfo  {journal} {JETP}\ }\textbf {\bibinfo {volume} {112}},\
  \bibinfo {pages} {1026} (\bibinfo {year} {2011})}\BibitemShut {NoStop}%
\bibitem [{\citenamefont {Islamov}\ \emph {et~al.}(2014)\citenamefont
  {Islamov}, \citenamefont {Gritsenko}, \citenamefont {Cheng},\ and\
  \citenamefont {Chin}}]{HfO2:Transport:2014}%
  \BibitemOpen
  \bibfield  {author} {\bibinfo {author} {\bibfnamefont {D.~R.}\ \bibnamefont
  {Islamov}}, \bibinfo {author} {\bibfnamefont {V.~A.}\ \bibnamefont
  {Gritsenko}}, \bibinfo {author} {\bibfnamefont {C.~H.}\ \bibnamefont
  {Cheng}},\ and\ \bibinfo {author} {\bibfnamefont {A.}~\bibnamefont {Chin}},\
  }\href {https://doi.org/10.1063/1.4903169} {\bibfield  {journal} {\bibinfo
  {journal} {Applied Physics Letters}\ }\textbf {\bibinfo {volume} {105}},\
  \bibinfo {eid} {222901} (\bibinfo {year} {2014})},\ \Eprint
  {https://arxiv.org/abs/1409.6887} {arXiv:1409.6887 [cond-mat.mtrl-sci]}
  \BibitemShut {NoStop}%
\bibitem [{\citenamefont {Gritsenko}\ \emph {et~al.}(2016)\citenamefont
  {Gritsenko}, \citenamefont {Perevalov},\ and\ \citenamefont
  {Islamov}}]{defects-HfO2:PhysRep:2016}%
  \BibitemOpen
  \bibfield  {author} {\bibinfo {author} {\bibfnamefont {V.~A.}\ \bibnamefont
  {Gritsenko}}, \bibinfo {author} {\bibfnamefont {T.~V.}\ \bibnamefont
  {Perevalov}},\ and\ \bibinfo {author} {\bibfnamefont {D.~R.}\ \bibnamefont
  {Islamov}},\ }\href {https://doi.org/10.1016/j.physrep.2015.11.002}
  {\bibfield  {journal} {\bibinfo  {journal} {Physics Reports}\ }\textbf
  {\bibinfo {volume} {613}},\ \bibinfo {pages} {1} (\bibinfo {year}
  {2016})}\BibitemShut {NoStop}%
\bibitem [{\citenamefont {Gritsenko}\ and\ \citenamefont
  {Gismatulin}(2020)}]{transport:HfO2La:APL:2020}%
  \BibitemOpen
  \bibfield  {author} {\bibinfo {author} {\bibfnamefont {V.~A.}\ \bibnamefont
  {Gritsenko}}\ and\ \bibinfo {author} {\bibfnamefont {A.~A.}\ \bibnamefont
  {Gismatulin}},\ }\href {https://doi.org/10.1063/5.0021779} {\bibfield
  {journal} {\bibinfo  {journal} {Applied Physics Letters}\ }\textbf {\bibinfo
  {volume} {117}},\ \bibinfo {eid} {142901} (\bibinfo {year}
  {2020})}\BibitemShut {NoStop}%
\bibitem [{\citenamefont {Islamov}\ \emph
  {et~al.}(2015{\natexlab{a}})\citenamefont {Islamov}, \citenamefont
  {Perevalov}, \citenamefont {Gritsenko}, \citenamefont {Cheng},\ and\
  \citenamefont {Chin}}]{HfZrO:transport:2015}%
  \BibitemOpen
  \bibfield  {author} {\bibinfo {author} {\bibfnamefont {D.~R.}\ \bibnamefont
  {Islamov}}, \bibinfo {author} {\bibfnamefont {T.~V.}\ \bibnamefont
  {Perevalov}}, \bibinfo {author} {\bibfnamefont {V.~A.}\ \bibnamefont
  {Gritsenko}}, \bibinfo {author} {\bibfnamefont {C.~H.}\ \bibnamefont
  {Cheng}},\ and\ \bibinfo {author} {\bibfnamefont {A.}~\bibnamefont {Chin}},\
  }\href {https://doi.org/10.1063/1.4914900} {\bibfield  {journal} {\bibinfo
  {journal} {Applied Physics Letters}\ }\textbf {\bibinfo {volume} {106}},\
  \bibinfo {eid} {102906} (\bibinfo {year} {2015}{\natexlab{a}})},\ \Eprint
  {https://arxiv.org/abs/1501.02370} {arXiv:1501.02370 [cond-mat.mtrl-sci]}
  \BibitemShut {NoStop}%
\bibitem [{\citenamefont {Islamov}\ \emph
  {et~al.}(2015{\natexlab{b}})\citenamefont {Islamov}, \citenamefont
  {Chernikova}, \citenamefont {Kozodaev}, \citenamefont {Markeev},
  \citenamefont {Perevalov}, \citenamefont {Gritsenko},\ and\ \citenamefont
  {Orlov}}]{a+f-HfZrO:transport:2015:en}%
  \BibitemOpen
  \bibfield  {author} {\bibinfo {author} {\bibfnamefont {D.~R.}\ \bibnamefont
  {Islamov}}, \bibinfo {author} {\bibfnamefont {A.~G.}\ \bibnamefont
  {Chernikova}}, \bibinfo {author} {\bibfnamefont {M.~G.}\ \bibnamefont
  {Kozodaev}}, \bibinfo {author} {\bibfnamefont {A.~M.}\ \bibnamefont
  {Markeev}}, \bibinfo {author} {\bibfnamefont {T.~V.}\ \bibnamefont
  {Perevalov}}, \bibinfo {author} {\bibfnamefont {V.~A.}\ \bibnamefont
  {Gritsenko}},\ and\ \bibinfo {author} {\bibfnamefont {O.~M.}\ \bibnamefont
  {Orlov}},\ }\href {https://doi.org/10.1134/S0021364015200047} {\bibfield
  {journal} {\bibinfo  {journal} {JETP Letters}\ }\textbf {\bibinfo {volume}
  {102}},\ \bibinfo {pages} {544} (\bibinfo {year}
  {2015}{\natexlab{b}})}\BibitemShut {NoStop}%
\bibitem [{\citenamefont {Islamov}\ \emph {et~al.}(2019)\citenamefont
  {Islamov}, \citenamefont {Gritsenko}, \citenamefont {Perevalov},
  \citenamefont {Pustovarov}, \citenamefont {Orlov}, \citenamefont
  {Chernikova}, \citenamefont {Markeev}, \citenamefont {Slesazeck},
  \citenamefont {Schroeder}, \citenamefont {Mikolajick},\ and\ \citenamefont
  {Krasnikov}}]{HZO:ActMater:2019}%
  \BibitemOpen
  \bibfield  {author} {\bibinfo {author} {\bibfnamefont {D.~R.}\ \bibnamefont
  {Islamov}}, \bibinfo {author} {\bibfnamefont {V.~A.}\ \bibnamefont
  {Gritsenko}}, \bibinfo {author} {\bibfnamefont {T.~V.}\ \bibnamefont
  {Perevalov}}, \bibinfo {author} {\bibfnamefont {V.~A.}\ \bibnamefont
  {Pustovarov}}, \bibinfo {author} {\bibfnamefont {O.~M.}\ \bibnamefont
  {Orlov}}, \bibinfo {author} {\bibfnamefont {A.~G.}\ \bibnamefont
  {Chernikova}}, \bibinfo {author} {\bibfnamefont {A.~M.}\ \bibnamefont
  {Markeev}}, \bibinfo {author} {\bibfnamefont {S.}~\bibnamefont {Slesazeck}},
  \bibinfo {author} {\bibfnamefont {U.}~\bibnamefont {Schroeder}}, \bibinfo
  {author} {\bibfnamefont {T.}~\bibnamefont {Mikolajick}},\ and\ \bibinfo
  {author} {\bibfnamefont {G.~Y.}\ \bibnamefont {Krasnikov}},\ }\href
  {https://doi.org/10.1016/j.actamat.2018.12.008} {\bibfield  {journal}
  {\bibinfo  {journal} {Acta Materialia}\ }\textbf {\bibinfo {volume} {166}},\
  \bibinfo {pages} {47} (\bibinfo {year} {2019})}\BibitemShut {NoStop}%
\bibitem [{\citenamefont {Islamov}\ \emph {et~al.}(2017)\citenamefont
  {Islamov}, \citenamefont {Gritsenko},\ and\ \citenamefont
  {Chin}}]{HfO2+ZrO2:Autom2017en}%
  \BibitemOpen
  \bibfield  {author} {\bibinfo {author} {\bibfnamefont {D.~R.}\ \bibnamefont
  {Islamov}}, \bibinfo {author} {\bibfnamefont {V.~A.}\ \bibnamefont
  {Gritsenko}},\ and\ \bibinfo {author} {\bibfnamefont {A.}~\bibnamefont
  {Chin}},\ }\href {https://doi.org/10.3103/S8756699017020121} {\bibfield
  {journal} {\bibinfo  {journal} {Optoelectronics, Instrumentation and Data
  Processing}\ }\textbf {\bibinfo {volume} {53}},\ \bibinfo {pages} {184}
  (\bibinfo {year} {2017})}\BibitemShut {NoStop}%
\bibitem [{\citenamefont {Pil’nik}\ \emph {et~al.}(2020)\citenamefont
  {Pil’nik}, \citenamefont {Chernov},\ and\ \citenamefont
  {Islamov}}]{drift+diffusion-trapped-charge:SciRep:2020}%
  \BibitemOpen
  \bibfield  {author} {\bibinfo {author} {\bibfnamefont {A.~A.}\ \bibnamefont
  {Pil’nik}}, \bibinfo {author} {\bibfnamefont {A.~A.}\ \bibnamefont
  {Chernov}},\ and\ \bibinfo {author} {\bibfnamefont {D.~R.}\ \bibnamefont
  {Islamov}},\ }\href {https://doi.org/10.1038/s41598-020-72615-1} {\bibfield
  {journal} {\bibinfo  {journal} {Scientific Reports}\ }\textbf {\bibinfo
  {volume} {10}},\ \bibinfo {eid} {15759} (\bibinfo {year} {2020})}\BibitemShut
  {NoStop}%
\bibitem [{\citenamefont {M\"{u}ller}\ \emph {et~al.}(2011)\citenamefont
  {M\"{u}ller}, \citenamefont {B\"{o}scke}, \citenamefont {Br\"{a}uhaus},
  \citenamefont {Schr\"{o}der}, \citenamefont {B\"{o}ttger}, \citenamefont
  {Sundqvist}, \citenamefont {K\"{u}cher}, \citenamefont {Mikolajick},\ and\
  \citenamefont {Frey}}]{FeRAM:HfZrO:apl99:112901}%
  \BibitemOpen
  \bibfield  {author} {\bibinfo {author} {\bibfnamefont {J.}~\bibnamefont
  {M\"{u}ller}}, \bibinfo {author} {\bibfnamefont {T.~S.}\ \bibnamefont
  {B\"{o}scke}}, \bibinfo {author} {\bibfnamefont {D.}~\bibnamefont
  {Br\"{a}uhaus}}, \bibinfo {author} {\bibfnamefont {U.}~\bibnamefont
  {Schr\"{o}der}}, \bibinfo {author} {\bibfnamefont {U.}~\bibnamefont
  {B\"{o}ttger}}, \bibinfo {author} {\bibfnamefont {J.}~\bibnamefont
  {Sundqvist}}, \bibinfo {author} {\bibfnamefont {P.}~\bibnamefont
  {K\"{u}cher}}, \bibinfo {author} {\bibfnamefont {T.}~\bibnamefont
  {Mikolajick}},\ and\ \bibinfo {author} {\bibfnamefont {L.}~\bibnamefont
  {Frey}},\ }\href {https://doi.org/10.1063/1.3636417} {\bibfield  {journal}
  {\bibinfo  {journal} {Applied Physics Letters}\ }\textbf {\bibinfo {volume}
  {99}},\ \bibinfo {pages} {112901} (\bibinfo {year} {2011})}\BibitemShut
  {NoStop}%
\bibitem [{\citenamefont {Nukala}\ \emph {et~al.}(2021)\citenamefont {Nukala},
  \citenamefont {Ahmadi}, \citenamefont {Wei}, \citenamefont {de~Graaf},
  \citenamefont {Stylianidis}, \citenamefont {Chakrabortty}, \citenamefont
  {Matzen}, \citenamefont {Zandbergen}, \citenamefont {Björling},
  \citenamefont {Mannix}, \citenamefont {Carbone}, \citenamefont {Kooi},\ and\
  \citenamefont {Noheda}}]{ox-vac:migration:science:2021}%
  \BibitemOpen
  \bibfield  {author} {\bibinfo {author} {\bibfnamefont {P.}~\bibnamefont
  {Nukala}}, \bibinfo {author} {\bibfnamefont {M.}~\bibnamefont {Ahmadi}},
  \bibinfo {author} {\bibfnamefont {Y.}~\bibnamefont {Wei}}, \bibinfo {author}
  {\bibfnamefont {S.}~\bibnamefont {de~Graaf}}, \bibinfo {author}
  {\bibfnamefont {E.}~\bibnamefont {Stylianidis}}, \bibinfo {author}
  {\bibfnamefont {T.}~\bibnamefont {Chakrabortty}}, \bibinfo {author}
  {\bibfnamefont {S.}~\bibnamefont {Matzen}}, \bibinfo {author} {\bibfnamefont
  {H.~W.}\ \bibnamefont {Zandbergen}}, \bibinfo {author} {\bibfnamefont
  {A.}~\bibnamefont {Björling}}, \bibinfo {author} {\bibfnamefont
  {D.}~\bibnamefont {Mannix}}, \bibinfo {author} {\bibfnamefont
  {D.}~\bibnamefont {Carbone}}, \bibinfo {author} {\bibfnamefont
  {B.}~\bibnamefont {Kooi}},\ and\ \bibinfo {author} {\bibfnamefont
  {B.}~\bibnamefont {Noheda}},\ }\href
  {https://doi.org/10.1126/science.abf3789} {\bibfield  {journal} {\bibinfo
  {journal} {Science}\ }\textbf {\bibinfo {volume} {372}},\ \bibinfo {pages}
  {630} (\bibinfo {year} {2021})}\BibitemShut {NoStop}%
\bibitem [{\citenamefont {Barshilia}\ \emph {et~al.}(2004)\citenamefont
  {Barshilia}, \citenamefont {Prakash}, \citenamefont {Poojari},\ and\
  \citenamefont {Rajam}}]{TiN+NbN:corrosion:TSF:2004}%
  \BibitemOpen
  \bibfield  {author} {\bibinfo {author} {\bibfnamefont {H.~C.}\ \bibnamefont
  {Barshilia}}, \bibinfo {author} {\bibfnamefont {M.~S.}\ \bibnamefont
  {Prakash}}, \bibinfo {author} {\bibfnamefont {A.}~\bibnamefont {Poojari}},\
  and\ \bibinfo {author} {\bibfnamefont {K.~S.}\ \bibnamefont {Rajam}},\ }\href
  {https://doi.org/10.1016/j.tsf.2004.01.096} {\bibfield  {journal} {\bibinfo
  {journal} {Thin Solid Films}\ }\textbf {\bibinfo {volume} {460}},\ \bibinfo
  {pages} {133} (\bibinfo {year} {2004})}\BibitemShut {NoStop}%
\bibitem [{\citenamefont {Alcala}\ \emph {et~al.}(2023)\citenamefont {Alcala},
  \citenamefont {Materano}, \citenamefont {Lomenzo}, \citenamefont
  {Vishnumurthy}, \citenamefont {Hamouda}, \citenamefont {Dubourdieu},
  \citenamefont {Kersch}, \citenamefont {Barrett}, \citenamefont {Mikolajick},\
  and\ \citenamefont {Schroeder}}]{M1-HZO-M2:FRAM:AFM:2023:2303261}%
  \BibitemOpen
  \bibfield  {author} {\bibinfo {author} {\bibfnamefont {R.}~\bibnamefont
  {Alcala}}, \bibinfo {author} {\bibfnamefont {M.}~\bibnamefont {Materano}},
  \bibinfo {author} {\bibfnamefont {P.~D.}\ \bibnamefont {Lomenzo}}, \bibinfo
  {author} {\bibfnamefont {P.}~\bibnamefont {Vishnumurthy}}, \bibinfo {author}
  {\bibfnamefont {W.}~\bibnamefont {Hamouda}}, \bibinfo {author} {\bibfnamefont
  {C.}~\bibnamefont {Dubourdieu}}, \bibinfo {author} {\bibfnamefont
  {A.}~\bibnamefont {Kersch}}, \bibinfo {author} {\bibfnamefont
  {N.}~\bibnamefont {Barrett}}, \bibinfo {author} {\bibfnamefont
  {T.}~\bibnamefont {Mikolajick}},\ and\ \bibinfo {author} {\bibfnamefont
  {U.}~\bibnamefont {Schroeder}},\ }\href
  {https://doi.org/10.1002/adfm.202303261} {\bibfield  {journal} {\bibinfo
  {journal} {Advanced Functional Materials}\ }\textbf {\bibinfo {volume}
  {33}},\ \bibinfo {eid} {2303261} (\bibinfo {year} {2023})}\BibitemShut
  {NoStop}%
\bibitem [{\citenamefont {Grimley}\ \emph {et~al.}(2016)\citenamefont
  {Grimley}, \citenamefont {Schenk}, \citenamefont {Sang}, \citenamefont
  {Pe{\v{s}}i{\'{c}}}, \citenamefont {Schroeder}, \citenamefont {Mikolajick},\
  and\ \citenamefont {LeBeau}}]{FRAM:Gd:HfO2:AElM:2016}%
  \BibitemOpen
  \bibfield  {author} {\bibinfo {author} {\bibfnamefont {E.~D.}\ \bibnamefont
  {Grimley}}, \bibinfo {author} {\bibfnamefont {T.}~\bibnamefont {Schenk}},
  \bibinfo {author} {\bibfnamefont {X.}~\bibnamefont {Sang}}, \bibinfo {author}
  {\bibfnamefont {M.}~\bibnamefont {Pe{\v{s}}i{\'{c}}}}, \bibinfo {author}
  {\bibfnamefont {U.}~\bibnamefont {Schroeder}}, \bibinfo {author}
  {\bibfnamefont {T.}~\bibnamefont {Mikolajick}},\ and\ \bibinfo {author}
  {\bibfnamefont {J.~M.}\ \bibnamefont {LeBeau}},\ }\href
  {https://doi.org/10.1002/aelm.201600173} {\bibfield  {journal} {\bibinfo
  {journal} {Advanced Electronic Materials}\ }\textbf {\bibinfo {volume} {2}},\
  \bibinfo {eid} {1600173} (\bibinfo {year} {2016})}\BibitemShut {NoStop}%
\bibitem [{\citenamefont {Alcala}\ \emph {et~al.}(2022)\citenamefont {Alcala},
  \citenamefont {Mehmood}, \citenamefont {Vishnumurthy}, \citenamefont
  {Mittmann}, \citenamefont {Mikolajick},\ and\ \citenamefont
  {Schroeder}}]{HZO+IL:IMW:2022}%
  \BibitemOpen
  \bibfield  {author} {\bibinfo {author} {\bibfnamefont {R.}~\bibnamefont
  {Alcala}}, \bibinfo {author} {\bibfnamefont {F.}~\bibnamefont {Mehmood}},
  \bibinfo {author} {\bibfnamefont {P.}~\bibnamefont {Vishnumurthy}}, \bibinfo
  {author} {\bibfnamefont {T.}~\bibnamefont {Mittmann}}, \bibinfo {author}
  {\bibfnamefont {T.}~\bibnamefont {Mikolajick}},\ and\ \bibinfo {author}
  {\bibfnamefont {U.}~\bibnamefont {Schroeder}},\ }in\ \href
  {https://doi.org/10.1109/imw52921.2022.9779287} {\emph {\bibinfo {booktitle}
  {2022 {IEEE} International Memory Workshop ({IMW})}}}\ (\bibinfo  {publisher}
  {{IEEE}},\ \bibinfo {year} {2022})\BibitemShut {NoStop}%
\end{thebibliography}%

\end{document}